\documentclass[fleqn,usenatbib]{mnras}

\usepackage{newtxtext,newtxmath}

\usepackage[T1]{fontenc}
\usepackage{siunitx}

\DeclareRobustCommand{\VAN}[3]{#2}
\let\VANthebibliography\thebibliography
\def\thebibliography{\DeclareRobustCommand{\VAN}[3]{##3}\VANthebibliography}
\newcommand{\MSun}{{\rm M}_\odot}

\usepackage{graphicx}	% Including figure files
\usepackage{amsmath}	% Advanced maths commands
\usepackage{amssymb}	% Extra maths symbols

\title[Social distancing in the Universe]{Social distancing between particles and objects in the Universe}

\author[Flonkelaar et al.]{
Door van Flonkelaar$^{1}$, Bozef Jucko$^{1}$, Gudit Marg$^{1}$, Koah Nubli$^{1}$, Schebastian Sulz$^{1}$
\\
$^1$Center for Theoretical Astrophysics and Cosmology, Institute for Computational Science, University of Zurich, Winterthurerstrasse 190, CH-8057 Z\"urich, Switzerland 
}

\date{Accepted today. Received today; in original form today}

\pubyear{2022}

\begin{document}
\label{firstpage}
\pagerange{\pageref{firstpage}--\pageref{lastpage}}
\maketitle

\begin{abstract}
The novel coronavirus, dubbed COVID-19, upended our lives may be in irreversible ways during its initial spread throughout the world in March 2020. It forced us all, willingly or unwillingly, to keep social distance from each other to slow down the spread of COVID-19. As scientists, we started speculating what kind of separation is between the constitutes of different objects in the Universe. In this work, we study the "social" distance between elements inside various objects, no matter their size, mass, and nature. We consider things ranging from diamond, baseball to Saturn, asteroid belt or M87 Black Hole, to name a few. We show our results in the form of a fascinating mass/"social" distance plot, where a cool cartoon figure represents each object. 
\end{abstract}


%%%%%%%%%%%%%%%%%%%%%%%%%%%%%%%%%%%%%%%%%%%%%%%%%%

%%%%%%%%%%%%%%%%% BODY OF PAPER %%%%%%%%%%%%%%%%%%

\section{Introduction}
China reported a novel coronavirus (COVID-19) in Wuhan, Hubei Province, on December 31, 2019 \citep{gralinski2020return}. The COVID-19 pandemic is considered the greatest public health threat since the 1918 Influenza Pandemic that infected one-third of the world’s population and killed at least 50 million people. COVID-19 cases and fatalities were and are still growing exponentially and there is much uncertainty about its ultimate impacts globally \citep{bib}. One of the key questions of public health epidemiology is how individual actions can help mitigate and manage the costs of such a pandemic \citep{reluga2010game}. 

Social distancing is an aspect of human behavior particularly important to epidemiology because of its universality; everybody can reduce their contact rates by changing their behaviors. Theoretical work on social distancing has been stimulated by studies that indicate that small changes in behavior can have large effects on transmission patterns during an epidemic \citep{del2005effects}. 

Astronomers study, among other things, (most often large) objects and their meaningfull properties. The bulk of the  job is heavily based on research, as the focus is on understanding how the universe works, and on trying to discover things about the universe that would be considered scientific breakthroughs \citep{careerexplorer_2019}. To help the world during these difficult times, we visualized the social distancing between particles and object in the Universe to get a better view on how far 1.5 meter actually is between two humans. The final result can be found in figure \ref{fig:the_plot}.

This paper is organized as follows. In Section 2 we describe the method and calculations describing how we find the values for the object mass and the social distancing property. In Section 3 we present our conclusion.   Finally we list the physical constants in Appendix A and our figure in Appendix B.

\section{Methods}
Below we describe how we determined the mass and distance between the objects. 

\subsection{Objects on earth}
\label{sec:lab} % used for referring to this section from elsewhere
We considered 5 different objects one can find on earth: water, uranium, a baseball and  a diamond. For this objects we take the weight of 1 dm$^3$, or in case of the baseball and diamond\footnote{0.6 carat} the average weight of the object. this gives us the values shown in table \ref{objects:tab}. To determine the molecular distance between the particles we find the number of molecules in 1 dm$^3$ and use $D = \frac{1}{\sqrt[3]{N}}$, where N is the number density. The distance between the molecules is given in table \ref{objects:tab}.

\begin{table}
\begin{tabular}{lcccc}
\hline
 & \multicolumn{1}{l}{\textbf{Mass (kg)}} & \textbf{u} & \textbf{N} & \textbf{Distance (m)} \\ \hline
\textbf{Water}    & 1     & 18.02  & 3.34E25 & 3.10E-10 \\ \hline
\textbf{Uranium}  & 19.05 & 238.03 & 4.82E25 & 2.75E-10 \\ \hline
\textbf{Baseball} & 0.15   & 68.12  &6.19E24 &5.45E-10\\ \hline
\textbf{Diamond}  & 0.00012  & 12.01 &1.76E26 &1.78E-10
\end{tabular}
\caption{The table shows the mass in kilogram, molecular weight in u, amount of molecules in 1 dm$^3$ and the distance between the molecules in m. For the baseball we took molecular weight of Polyisoprene \citep{Jones_Harris_1992}.}
\label{objects:tab}
\end{table}

\subsection{Humans}
\label{sec:human} % used for referring to this section from elsewhere

The average weight of an adult in Europe is 70.8 kg and Asia is 57.7 kg \citep[data from 2005,][]{weight2012}, for the population in Sweden and COVID-restricted we use the European average weight and for the population in Hong Kong we'll use the Asian average weight. 

In table \ref{tab:human}, one can see the population, surface area and average distance between inhabitants of Sweden and Hong Kong. For the distance we use $D = \frac{1}{\sqrt{N}}$, where N is the population density. Data is from \citet{WB:2021}. In Switzerland a social distancing of 1.5 metres is maintained during COVID times \citep{bag:2021}.
\begin{table}
\begin{tabular}{lccc}
\hline
                   & \textbf{Population} & \textbf{Area (km$^2$)} & \textbf{Distance (m)} \\ \hline
\textbf{Sweden}    & 10,350,000          & 450,295                & 660.87                    \\ \hline
\textbf{Hong kong} & 7,482,000           & 1,106                  & 12.16                     \\ \hline
\end{tabular}
\caption{The population, surface area and average distance between inhabitants of Sweden and Hong Kong \citep{WB:2021}.}
\label{tab:human}
\end{table}

\subsection{M87 black hole} 
M = $m_p$ \& r = $\left(\frac{32\pi G^3(6.5\times10^9\MSun)^2m_p}{3c^6}\right)^{\frac13}$

Supermassive Black hole with approximate mass  $M87\approx6.5x10^9~\MSun$ at the centre of Messier 87 or M87 elliptical Galaxy at a distance of around 54 million ly \citep{EHT2019}. 

If take the particle mass $m_p$ then we can calculate separation using density ($\rho$) estimate inside event horizon radius $\frac{2GM}{c^2}$  
\begin{equation}
    \rho_\text{M87}=\frac{M_\text{M87}}{\frac{4\pi}{3}\left(\frac{2GM_\text{M87}}{c^2}\right)^3}=\frac{m_p}{r^3}\implies
    r=\left(\frac{32\pi G^3M_\text{M87}^2m_p}{3c^6}\right)^{\frac13}
\end{equation}

\subsection{Neutron star}
M = $m_n$ \& r = $\frac{h^2} {(1.4G\MSun)^{\frac23}(Gm_n)^{\frac13}m_n^2}$

Formed at the end of star's life as its collapsed core. Mass ranges from one to three $\MSun$. Mostly made up of neutrons, it could also have electrons, protons, muons, etc. in relatively small quantity.

Approximate radius for relativistic particles dominated object is 
\begin{equation}
    R\sim\frac{h^2}{Gm_Em^{\frac53}M_\text{tot}^{\frac13}},
\end{equation}
where $m_E$ is the particle which carry energy, $m$ is the particle mass which dominates the total mass, and $M$ is the total mass. Hence if we have M=$m_n$ then separation has to be
\begin{equation}\label{Eq:Separation}
    r\sim\frac{h^2}{(GM_\text{tot})^{\frac23}m_Em(Gm)^{\frac13}(m m_n^{-1})^{\frac13}}.
\end{equation}
For neutron star, we have $M_\text{tot}=1.4\MSun$, $m_E=m=m_n$ which gives us given separation.

\subsection{Brown dwarf} 
M = $m_n$ \& r = $\frac{h^2} {(0.075 G\MSun)^{\frac23}(Gm_n)^{\frac13}m_em_n}$

Substellar object which are massive than planets but lighter than stars. Typically 0.075 times lighter than the Sun with lowest possible mass around 13 Jupiter mass. Brown dwarf unlike stars do not sustain in their core nuclear fusion of hydrogen into helium hence resulting in unstable luminosities.

Plugging $M_\text{tot}=0.075\MSun$, $m_E=m_e,~m=m_n$ into equation \eqref{Eq:Separation}, we get the given separation.

\subsection{Nebula}
M = $m_p$ \& r = $10^{-3}m$

Made up of interstellar clouds. Possible remnant of a dying star or the location where stars are formed. Typically have 100s of ly diameter and Orion Nebula could be seen with naked eyes. 

If we assume purely hydrogen Nebula with typical particle density of 1000 $cm^{-3}$ then if $M=m_p$ then separation is $r=10^{-3}$. 

\subsection{Nuclear star cluster}
M = $\MSun$ \& r = 0.03pc

Massive compact star cluster with high density and high luminosity found near the centre of most galaxies. Dwarf galaxies with no central supermassive black hole usually host nuclear star cluster at their centre.

If we have a nuclear star cluster with total mass of 10$^6\MSun$ and effective radius of 3 pc hence for $M=\MSun$ we have $r=0.03$pc.

\subsection{Locust swarm}
M =  0.002 Kg \& r = 0.13 m

Swarm of short-horned grasshoppers, which are initially innocuous solitary beings later turned into swarmy winged gregarious insects under suitable conditions. In this later phase, they start destroying the agricultural entities.

If we assume on an average mass of 2g per locust then for locust swarm density of 60 million per km$^2$ we get r = 0.13 m.

\subsection{Solar photons on Earth}

\begin{equation}
    I_{\odot} = \int_{0}^{\infty} B(\lambda,T=T_\odot) {\rm d}\lambda, 
\end{equation}
with $T_\odot = 5777 \rm{K}$ and
\begin{equation}
    B(\lambda,T) = \frac{2hc^2}{\lambda^5}\frac{1}{e^{\frac{hc}{\lambda k_B T}}-1}
\end{equation}
Solar luminosity $L_\odot = I_{\odot} (4 \pi 1 \rm{AU})^2 \doteq 4 \cdot 10^{26} \,\rm W/s$.

If we divide the solar radiation intensity $I_\odot$ by the energy of solar photons $E = hc/\lambda_\odot$ we get the number density flux. By further division by speed of light, we obtain number density

\begin{equation}
    n_\odot = \frac{\lambda L_\odot}{4\pi  hc^2 (1 \rm{AU})^2} = 1.2\cdot 10^{13}\, \rm{m^{-3}}
\end{equation}
and therefore the interparticle distance can be achieved
\begin{equation}
    l_\odot = n_\odot^{-1/3} = 5\cdot 10^{-5} \, \rm m.
\end{equation}

Although the photons are massless particles, we can assign a mass to them based on their energy as
\begin{equation}
    \MSun^\gamma = \frac{h}{\lambda_\odot c} = 4\cdot 10^{-36}\,\rm kg 
\end{equation}

\subsection{CMB photons}

Cosmic microwave background (CMB), a relativistic species, has number density which can be calculated using Bose-Einstein phase space distribution function $f(\mathbf{p}) = \left(e^{\frac{E(\mathbf{p})}{k_B T}} - 1\right)^{-1}$: 

\begin{equation}
    n_{\rm eq}(\mathbf{p}) = \frac{g_\gamma}{(2\pi)^3} \int f(\mathbf{p}) \rm{d}\mathbf{p},
\end{equation}
where $g_\gamma = 2$ for photons. By integrating, one gets

\begin{equation}
    n_{\rm eq} = \frac{\zeta(3)}{\pi^2} g_\gamma \left( \frac{k_B T}{h c}\right)^3.
\end{equation}
After plugging in the specific values, namely $T=T_{\rm CMB} = 2.73 \rm K$, we get $n_{\rm eq} = 1.65\cdot 10^6\, \rm m^{-3}$. From number density, it is easy to obtain the average social distance between specific photons

\begin{equation}
    l_{\rm CMB} = n_{\rm eq}^{-1/3} = 8\cdot 10^{-3} \, \rm m.  
\end{equation}
Similar to the previous calculation, we can get the mass as  
\begin{equation}
    M^\gamma_{\rm CMB} = \frac{h}{\lambda_{\rm CMB}c} = 1.1\cdot 10^{-39}\,\rm kg 
\end{equation}
with $\lambda_{\rm CMB} \doteq 2\,\rm mm$.

\subsection{Cold atoms}

Here, we refer to the experiments with cold atoms in optical lattices. The social distance of cold atoms is given by the half-wavelength of laser generating periodic potential. We use data from \cite{cold_atoms}, where rubidium atoms $^{87}\rm Rb$ were trapped in potential generated by laser with wavelength $\lambda = 785.3\,\rm nm$. So the cold atoms social distance is $\lambda/2 = 3.9\cdot 10 ^{-7}$~m and the particle mass $m_{^{87}\rm Rb} \doteq 87 m_p = 1.4\cdot 10^{-25}$~kg.

\subsection{WIMP background}

We assume that all the dark matter content of the Universe is made of standard WIMPs (weakly interacting massive particles) with mass 100~GeV. Using the Planck 2018 results (\cite{Planck2018}), we take $\Omega_m = 0.315, \Omega_b = 0.049$. So $\Omega_{\rm dm} = \frac{\rho_{\rm dm}}{\rho_c} = \Omega_m - \Omega_b = 0.266$. With $\rho_c = \frac{3H^2}{8\pi G}$ and $H=67.37$~km/s/Mpc, the mass density of WIMP background can be estimated as 
\begin{align}
&\rho_{\rm dm} = \Omega_{\rm dm}\rho_c = 0.266\times 9\cdot 10^{-27} \rm{kg/m^{3}} =2.4\cdot 10^{-27} \rm{kg/m^{3}}\nonumber\\
&\rho_{\rm dm}c^2 \doteq 1.3 \rm GeV/m^3    
\end{align}
and therefore number density $n_{\rm WIMP} = 1.3\cdot 10^{-2} \rm m^{-3}$. As a result, the social distance between WIMP particles is $l_{\rm WIMP} \simeq 4.2~\rm m$.

\subsection{Giant planets/Saturn (core)}
We will pick Saturn as a representative of giant planets. According to \citep{Mankovich2021}, the Saturn's core extends as far as up to 60\% of the planet's radius. The cited study models the core density profile, reaching $\rho_{\rm core} = 7$~g/cm$^3$ at the centre of the planet and decreasing with larger radii. For the sake of this approximation, we take an average core density to be $\rho_{\rm core,av} = 3.5$~g/cm$^3$ (estimated from Fig.2b of \citep{Mankovich2021}). and assume the core composed mostly of iron. This gives a number density of $n = \rho_{\rm core,av}/56m_p \doteq 3.7 \cdot 10^{22}$~cm$^{-3}$. Mean social distance can be then easily obtained and gives $l_{\rm core} \doteq 3.0 \cdot 10^{-10}$~m.

\subsection{Oort cloud}
The quantitative description of Oort cloud is still very challenging and there is many uncertainties in the physical details of the system. In our work, we adopt the properties of  Oort cloud from \cite{stz2671}. We assume the Oort cloud to be a spherical system starting at approximately  $r_0 = 32$ AU measured from the centre of Solar system, where 1AU$\doteq1.5\cdot 10^{11}$ m and reaches as far as $R \sim 10^4$ AU. Further we assume, that the total number of particles it contains (planetesimals, comets, asteroids,...) is  $N=10^{11}$ and the total mass of the Oort cloud $M_{\rm total}\sim 10\rm{M}_\oplus$. In agreement with \cite{stz2671}, we assume the density profile of Oort cloud scaling as $\rho \sim r^{-3.5}$. In the case of spherical symmetry, for the total number of objects holds

\begin{equation}
    N = \rho_0\int_{r_0}^{R}\frac{1}{r^{3.5}}4 \pi r^2 \mathrm{d} r = 6\pi \rho_0\left[\frac{1}{\sqrt{r_0}} - \frac{1}{\sqrt{R}} \right].
\end{equation}
We as the Oort cloud size is significant and so is the density difference in the inner and outer edge, we will calculate the characteristic social distance in these two regions. For the inner region number density, one can write
\begin{equation}
    n_{\rm in} = \frac{\rho_0}{r_0^{3.5}M_{\rm OC}},
\end{equation}
where $M_{\rm OC} = \frac{M_{\rm total}}{N}$ is the average mass of the Oort cloud object. We assume that the average mass in the inner and outer region is similar. For social distance at $r = r_0$, we obtain

\begin{equation}
    l_{\rm in} = \frac{1}{n_{\rm in}^{1/3}} = r_0\left(\frac{6\pi}{N}\right)^{1/3}\left[1 - \sqrt{\frac{r_0}{R}} \right].
\end{equation}
Analogously, number density  in the outer region can be expressed as
\begin{equation}
    n_{\rm out} = \frac{\rho_0}{R^{3.5}M_{\rm OC}}
\end{equation}
and the average social distance then follows
\begin{equation}
    l_{\rm out} = \frac{1}{n_{\rm out}^{1/3}} = R\left(\frac{6\pi}{N}\right)^{1/3}\left[\sqrt{\frac{R}{r_0}} -1\right].
\end{equation}
Using the introduced values of Oort cloud properties, the social distances result into $l_{\rm in} = 2.7\cdot 10^9 \mathrm{ m} \sim 0.018 \,\mathrm{ AU}$ and  $l_{\rm out} = 2.2\cdot 10^{12} \mathrm{ m} \sim 15\, \mathrm{ AU}$. The average object mass $M_{\rm OC} = 6\cdot 10^{14}$~kg.

\subsection{Global IGM}
The global intergalactic medium (IGM) is the baryonic matter in the regions in between galaxies and galaxy clusters. It is made up mostly of ionized hydrogen \citep{Meiksin2009}, so the average mass of its constituent particles can be estimated as approximately one proton mass $m_\text{p}$. Taking the most recent results from Planck \citep{Planck2018} for the density parameter of baryonic matter in the universe, $\Omega_{b,0}$, together with the critical density of the universe, $\rho_\text{crit}$, we can estimate the average density of the baryonic matter, $\rho_\text{b,0}$, in the IGM.

\begin{equation}
    \frac{\rho_{\rm b,0}}{\rho_\text{crit}} = \Omega_{b,0} = 0.0493 \pm 0.035
\end{equation}

with

\begin{equation}
    \rho_\text{crit} = \frac{3 H_0^2}{8 \pi G}
\end{equation}

For the calculation of the critical density we will use the most recent result for the Hubble parameter from \citep{Planck2018}.

\begin{equation}
    H_0 = {67.36 \pm 0.54}\text{ km }\text{s}^{-1}\text{Mpc}^{-1}
\end{equation}

and hence the baryonic matter density is estimated as 

\begin{equation}
\rho_{\rm b,0} \approx 4.2 \cdot 10^{-28} \text{ kg }\rm m^{-3}
\end{equation}

which, assuming only protons, translates to 

\begin{equation}
    n_{\rm b, 0} \approx 2.5 \cdot 10^{-1} \text{ m}^{-3}
\end{equation}
Assuming the IGM particles are distributed evenly in a cubic volume, their average inter-particle distance is therefore
\begin{equation}
    d_\text{IGM} \approx 1.6 \text{ m}
\end{equation}

\subsection{Intra-cluster IGM}
The intra-cluster IGM is the baryonic matter that is in between galaxies in galaxy clusters. Similarly to the global IGM, it consists of mostly ionized hydrogen gas \citep{Sarazin1986}, so we estimate the average particle mass to be one proton mass $m_\text{p}$. The intra-cluster medium however is denser and hotter than the average global IGM as infalling material releases gravitational energy into the medium. Due to its high temperature, it emits strong X-ray radiation whose spectrum can be well explained by thermal bremsstrahlung \citep{Felten1966, Mushotzky1978}. Its average density can then be estimated from X-ray Luminosity measurements, which lead to a typical average intra-cluster IGM number density of approximately $10^{-3} \text{cm}^{-3}$ \citep{Felten1966, Sarazin1986, Loewenstein2003}. From the number density, is straightforward to estimate the social distance in a cubic volume of the intra-cluster IGM as $d_\text{intra} \approx 0.1\text{m}$, which is an order of magnitude less than that the one in the global IGM.

\subsection{ISM}
The interstellar medium is the baryonic matter that fills the space between the stars of the Milky Way galaxy. It has been studied thoroughly, and we direct the reader to a review on its properties by \citep{Ferriere2001} for a detailed summary. Weighted by mass, fifty percent of the ISM is in dense clouds of molecular gas, but these clouds only occupy one to two percent of its total volume. By volume, its main component is warm and mostly ionized hydrogen gas, so again we take the particle mass in the social setting of the ISM to be one proton mass. The average particle density can then be estimated by measuring the intensity of hydrogen Balmer emission lines that are produced after ionized hydrogen combines with a free electron to eventually transition from an excited state with $n > 2$ down to $n = 2$. These observations lead to an estimated average particle density of approximately $1 \text{cm}^{-3}$. Thus, in a cubic volume, the average social distance in the ISM is estimated to be $d_\text{ISM} \approx 10^{-2} \text{m}$.

\subsection{Dunlin swarm}
Some bird species move about collectively and occasionally gather in large swarms in order to feed, mate or to gain safety in numbers. It is an interesting question to ask what is the average interbird distance in such a swarm, and how does this distance relative to the individual bird's mass compare to the average "social" distance relative to the respective particle masses in the other systems treated in this paper. As a representative example, we estimate the social distance in a swarm of dunlins, which are small birds commonly found near shorelines of arctic and subarctic regions. A typical dunlin weighs $m_\text{dunlin} \approx 50 \text{g}$. Observations have shown that in a typical dunlin swarm of fifty birds, the average three-dimensional number density is approximately $4 \text{m}^{-3}$ \citep{Major2004}. This translates to an average interbird distance of $d_\text{dunlin} \approx 0.6 \text{m}$ in a cubic volume.

\subsection{Asteroid belt}
The asteroid belt is a torus-shaped region between Mars and Jupiter, ranging from $2 \text{AU}$ to $3.4 \text{AU}$, that contains a large amount of solid bodies which cover a wide range of sizes starting at diameters comparable to that of a dust particle and ending at the most massive object, Ceres, with a diameter of $950 \text{km}$. The total amount of mass in the asteroid belt is estimated to be $2.39 \cdot 10^{21} \text{kg}$ \citep{Pitjeva2018}, of which $\approx 60 \%$ are contributed from the four largest objects in it - Ceres, Vesta, Pallas and Hygiea. Most of the rest of the mass is then found in smaller objects with diameters larger than $1 \text{km}$ of which there are estimated to be between $7 \cdot 10^5$ and $1.7 \cdot 10^6$ \citep{Tedesco2002}. Therefore, on average, their mass is:

\begin{equation}
    m_\text{asteroid} \approx 2.39 \cdot 10^{21} \text{kg} \times 0.4/ 10^6 = 9.56 \cdot 10^{14} \text{kg} \approx 5 \cdot 10^{-16} \MSun
\end{equation}

Let us estimate the number density of these smaller objects in the asteroid belt by assuming they are equally distributed in an annulus between $2$ and $3.4 \text{AU}$ (In reality, the objects are neither equally distributed along the angles, nor do they share the same flat plane of orbits, but we believe our simplification suffices for an order of magnitude estimate of the social distance in the asteroid belt). The surface number density of objects is then:

\begin{equation}
    n = \frac{10^6} {\pi  \left(\left(3.4 \text{AU}\right)^2 - \left(2.0 \text{AU}\right)^2 \right)} \approx 1.87 \cdot 10^{-18} \text{m}^{-2}
\end{equation}

From this, we estimate the average distance between the objects to be $d_\text{belt} \approx 7.3 \cdot 10^8 \text{m}$.

\subsection{Air}
Assuming a density of $\rho = 10^{-3}\text{g}/\text{cm}^3$
and noting that the air consists mostly of Nitrogen molecules $N_2$,
having a mass of $28 m_p$, the average 
inter-particle distance yields:
$l = \sqrt[3]{\frac{3}{4\pi} \frac{28 m_p}{\rho}} \approx 2\cdot 10^{-7}\text{cm}$.
The number density gives:
$n = \frac{\rho}{28m_p} = 2.1\cdot 10^{19}\text{cm}^{-3}$

\subsection{Lab vacuum}
Assuming a vacuum of $10^{-6}\text{mbar}$ 
(Isolating vacuum of the LHC \citealt{noah_cern})
and applying the ideal gas equation the number density of molecules
yields: $n=P/k_bT = 2.4\cdot 10^{12} cm^{-3}$.
The inter-particle separation is
$l = \sqrt[3]{\frac{3}{4\pi n}} = 2.15\cdot 10^{-4}\text{cm}$

% \subsection{Stellar core (Noah)}
\subsection{Rocky planet}

Considering the most abundant elements in the earth, 
$32 \%$ Fe, $30 \%$ O, $15\%$ Si and $14 \%$ Mg
\cite{noah_planet},
one arrives at a mean molecular weight of $33 m_p$.
The density of the earth is $\rho = \frac{4\pi r_E^2}{3 M_E} = 5.5 \text{g} \text{cm}^{-3}$.
The inter-particle distance follows: 
\hbox{$l = \sqrt[3]{\frac{3}{4\pi} \frac{33 m_p}{\rho}}= 1.3\cdot 10^{-8}\text{cm}$}
and the particle density:
$n = \rho / (33 m_p) = 9.9\cdot 10^{23} \text{cm}^{-3}$.

\section{Conclusions}

As already argued, preserving sufficient interpersonal distance has been confirmed to be (among others) a very efficient precaution to suppress the spread of COVID-19 disease. Thus, we (to significant extent (hopefully)) lived in  a society with minimal "interparticle" (human) separation of $\sim$ 1.5 meters. In this work, we took a challenge of bringing up objects from the Earth as well as from the neighbouring or distant Universe across wide variety of scales in terms of masses and "social" (interparticle) distances together and provided a reader with a comprehensive overview of how such "social" distances can be calculated. Moreover, we visualized the chosen objects (credits to Door van Flonkelaar) and created an illustrative figure from which particle masses and "social" distances can be easily read out. In terms of distancing, we range from neutron stars with neutron separations of order $\sim 10^{-14}$~m to nuclear star clusters, in which stars are as far from each other as $\sim 10^{14}$~m.

From Fig.~\ref{fig:the_plot} we can observe that COVID-restricted humans, together with other terrestrial objects, are interestingly very close to the middle of the figure in terms of separation scales (however, one might argue adding even more exotic objects into the plot could modify this observation). During COVID-19, a separation of $~1.5$~m has shown to be a good compromise (however, some countries may have used 2 meters). However, in case of more severe circumstances, like a possible future Omega~-~variant,  we might be forced to keep larger mutual distances. If we assume the current population ($7.9$~billion) and the Earth surface area ($~510.1$~million km~$^2$), one can estimate a maximum interpersonal separation to be $\sim$~270~m (in case we could use 100\% of the Earth's surface). Whenever interpersonal distances of higher order of magnitude are needed, it's hardly imaginable to achieve that on Earth.

However looking at Fig.~\ref{fig:the_plot}, there is an optimism that, if need be in the future, there are systems in the Universe where such separations are feasible. We don't even have to look far, within our own solar system we could already do an "Arc of Noah" type mission by giving everyone their own Asteroid to stand on \citep{2020LPI....51.2430P}. 

Apart from that, this study does not provide any deeper arguments on how to approach a possibility of such a large "social" distancing or why such a review would practically help us in dealing with paradigm such as COVID-19, but it was fun. 

\section*{Acknowledgements}
The authors acknowledges Prof.\ Prasenjit Saha for insightful discussions and guidance.  Furthermore we would like to thank ICS for the existence of the pool room where some of the exciting ideas could emerge.

%%%%%%%%%%%%%%%%%%%%%%%%%%%%%%%%%%%%%%%%%%%%%%%%%%



\bibliographystyle{mnras}
\bibliography{Main} % if your bibtex file is called example.bib

%\begin{thebibliography}{99}
%\end{thebibliography}

%%%%%%%%%%%%%%%%%%%%%%%%%%%%%%%%%%%%%%%%%%%%%%%%%%

%%%%%%%%%%%%%%%%% APPENDICES %%%%%%%%%%%%%%%%%%%%%

\appendix

\section{Physical Constants}
\begin{align*}
    G&=6.67430(15)\times10^{-11}~\rm{m}^3~\rm{Kg}^{-1}~\rm{s}^{-2}~\text{(Gravitational Constant)}\\
    c&=299792458~\rm{m}~\rm{s}^{-1}~\text{(Speed of light in vacuum)}\\
    k_{\rm B}&=1.380649\times10^{-23}~\rm{m}^2~\rm{Kg}~\rm{s}^{-2}~\rm{K}^{-1}~\text{(Boltzmann constant)}\\
    h&=6.62607015\times10^{-34}~\rm{m}^2~\rm{Kg}~\rm{s}^{-1}~\text{(Planck's constant)}\\
    N_{\rm A}&=6.02214076\times10^{23}~\rm{mol^{-1}}~\text{(Avogadro constant)}\\
    m_{\rm e}&=9.1093837015(28)\times10^{-31}~\rm{Kg}~\text{(Mass of electron)}\\
    m_{\rm p}&= 1.672 621 923 69(51)\times10^{-27}~\rm{Kg}~\text{(Mass of proton)}\\
    m_{\rm n}&= 1.674 927 498 04(95)\times10^{-27}~\rm{Kg}~\text{(Mass of neutron)}\\
    \MSun&=1.989847(7)\times10^{30}~\rm{Kg}~\text{(Solar mass)}\\
    \rm{M}_\oplus&=5.9726(7)\times10^{24}~\rm{Kg}~\text{(Earth mass)}\\
    H_0&=67.36(54)\times\frac{10^5}{3.086\times10^{24}}~\rm{s}^{-1}~\text{(Hubble constant)}\\
    1 ly &=  9.4607\times10^{15}~\rm{m}~\text{(lightyear)}\\
    1 au &=  1.495978707\times10^{11}~\rm{m}~\text{(astronomical unit)}\\
    1 pc &=  3.0857\times10^{16}~\rm{m}~\text{(parsec)}\\
\end{align*}

\section{Figure}

%%%%%%%%%%%%%%%%%%%%%%%%%%%%%%%%%%%%%%%%%%%%%%%%%%

\begin{figure*}
    \rule{0cm}{5.5cm}
    \centering
    \includegraphics[width = \textwidth]{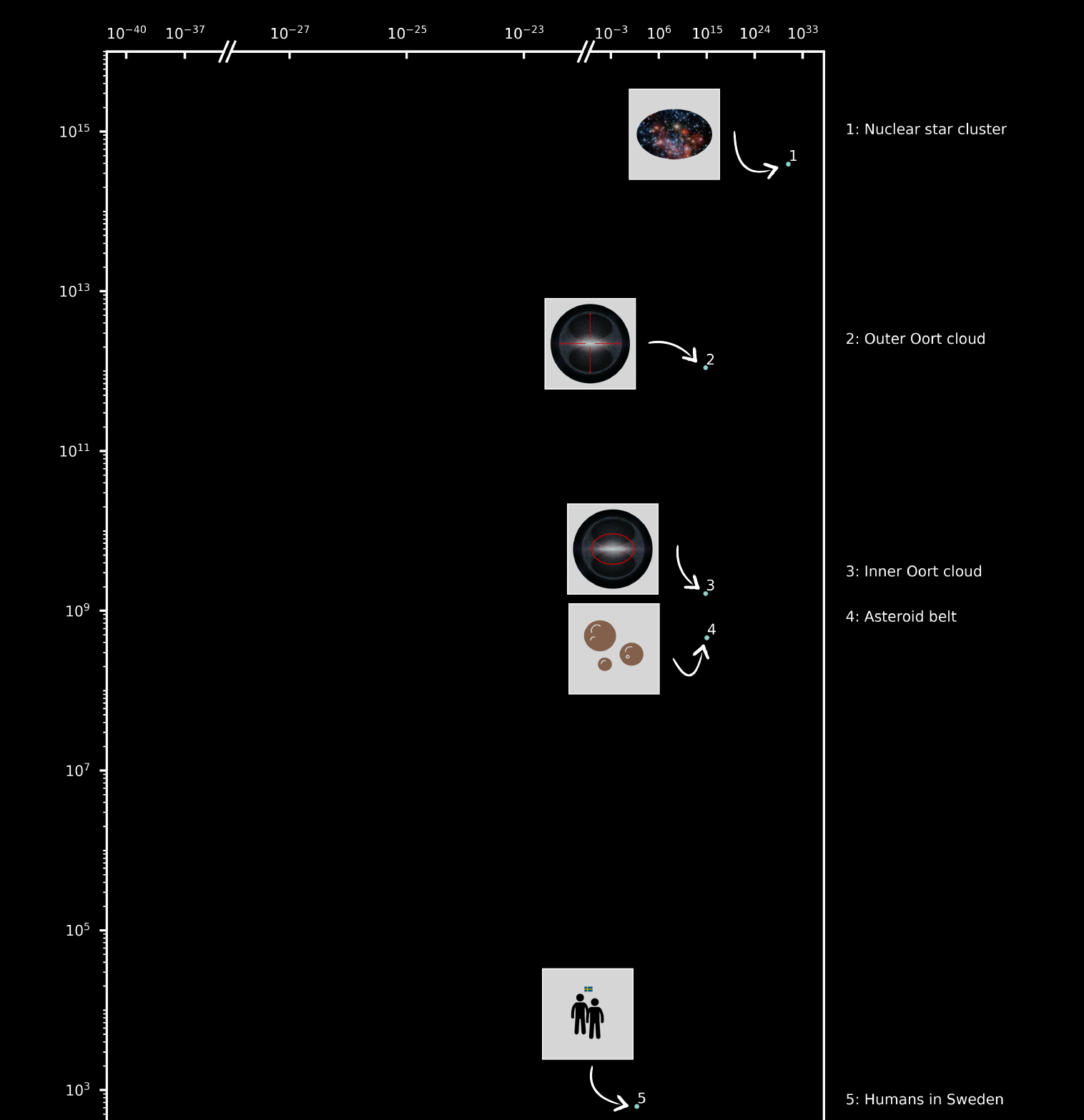}
    % \caption{THE plot}
\end{figure*}
\begin{figure*}
    \centering
    \includegraphics[width = \textwidth]{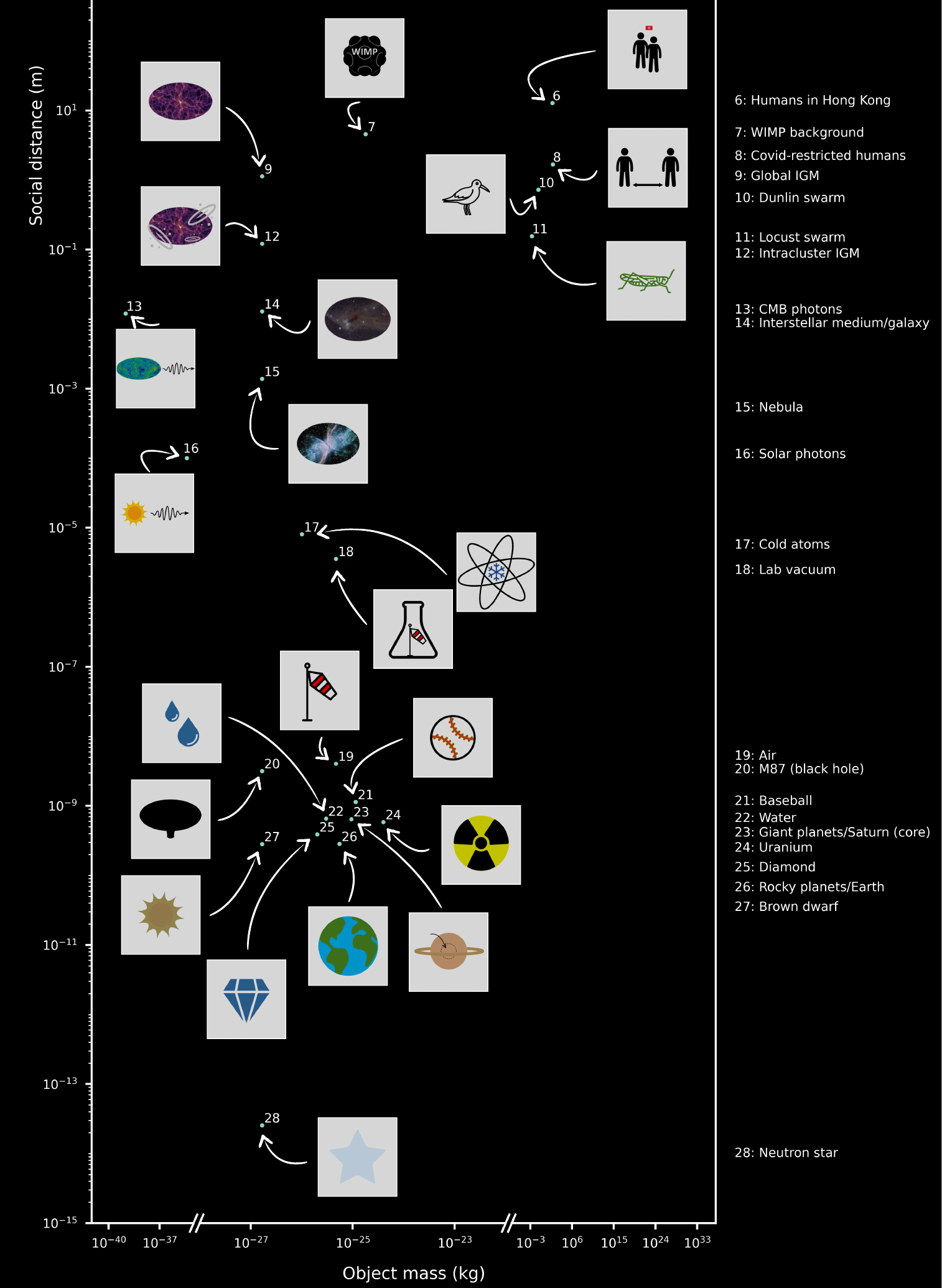}
    \caption{An illustration of different mass and social (interparticle) distance scales in a selection of  terrestrial, astrophysical and cosmological objects or media.}
    \label{fig:the_plot}
\end{figure*}

% Don't change these lines
\bsp	% typesetting comment

\label{lastpage}
\end{document}